\documentclass{aastex}
\usepackage{graphics}
\usepackage{graphicx}

\begin{document}

\title{The Search for Signatures Of Transient Mass Loss in Active Stars}

\author{M. K. Crosley}
\affil{Johns Hopkins University, Department of Physics \& Astronomy, 3400 N. Charles Street, Baltimore, MD 21218}

\author{R. A. Osten}
\affil{Space Telescope Science Institute, 3700 San Martin Dr, Baltimore, MD 21218}
\affil{Center for Astrophysical Sciences, Johns Hopkins University, Baltimore, MD 21218}

\author{J. W. Broderick}
\affil{ASTRON, the Netherlands Institute for Radio Astronomy, Postbus 2, 7990 AA Dwingeloo, The Netherlands}

\author{S. Corbel }
\affil{Laboratoire AIM (CEA/IRFU - CNRS/INSU - Universit\'e Paris Diderot), CEA DSM/IRFU/SAp, F-91191 Gif-sur-Yvette, France}
\affil{Station de Radioastronomie de Nan\c{c}ay, Observatoire de Paris, PSL Research University, CNRS, Univ. Orl\'{e}ans, OSUC, 18330 Nan\c{c}ay, France}

\author{J. Eisl\"{o}ffel }
\affil{Th\"{u}ringer Landessternwarte, Sternwarte 5, D-07778 Tautenburg, Germany}

\author{J.-M. Grie{\ss}meier}
\affil{LPC2E - Universit\'{e} d'Orl\'{e}ans / CNRS, 45071 Orl\'{e}ans cedex 2, France}
\affil{Station de Radioastronomie de Nan\c{c}ay, Observatorie de Paris, PSL Research University, CNRS, Univ. Orl\'{e}ans, OSUC, 18330 Nan\c{c}ay, France}

\author{J. van Leeuwen}
\affil{ASTRON, the Netherlands Institute for Radio Astronomy, Postbus 2, 7990 AA Dwingeloo, The Netherlands}

\author{A. Rowlinson }
\affil{Anton Pannekoek Institute for Astronomy, Science Park 904, 1098 XH Amsterdam, the Netherlands}
\affil{ASTRON, the Netherlands Institute for Radio Astronomy, Postbus 2, 7990 AA, Dwingeloo, The Netherlands}

\author{P. Zarka}
\affil{LESIA, Observatoire de Paris, CNRS, PSL/SU/UPMC/UPD/SPC, Place J. Janssen, 92195 Meudon, France}
\affil{Station de Radioastronomie de Nan\c{c}ay, Observatoire de Paris, PSL Research University, CNRS, Univ. Orl\'{e}ans, OSUC, 18330 Nan\c{c}ay, France}

\and

\author{C. Norman}
\affil{Johns Hopkins University, Department of Physics \& Astronomy, 3400 N. Charles Street, Baltimore, MD 21218}

\begin{abstract}

The habitability of an exoplanet depends on many factors.
One such factor is the impact of stellar eruptive events on nearby exoplanets.
Currently this is poorly constrained due to heavy reliance on solar scaling relationships and a lack of experimental evidence.  
Potential impacts of Coronal Mass Ejections (CMEs), which are a large eruption of magnetic field and plasma from a star, are  space weather and atmospheric stripping.
A method for observing CMEs as they travel though the stellar atmosphere is the type II radio burst, and the new LOw Frequency ARray (LOFAR) provides a means for detection.  
We report on 15 hours of observation of YZ Canis Minoris (YZ CMi), a nearby M dwarf flare star, taken in LOFAR's beam-formed observation mode for the purposes of measuring transient frequency-dependent low frequency radio emission. 
The observations utilized Low-Band Antenna (10-90 MHz) or High-Band Antenna (110-190 MHz) for five three-hour observation periods.  
In this data set, there were no confirmed type II events in this frequency range.  
We explore the range of parameter space for type II bursts constrained by our observations
Assuming the rate of shocks is a lower limit to the rate at which CMEs occur, no detections in a total of 15 hours of observation places a limit of $\nu_{type II} < 0.0667$ shocks/hr $ \leq \nu_{CME}$ for YZ CMi due to the stochastic nature of the events and limits of observational sensitivity.  
We propose a methodology to interpret jointly observed flares and CMEs which will provide greater constraints to CMEs and test the applicability of solar scaling relations.   

\end{abstract}


\section{Introduction}

M dwarfs are the most common type of star \citep{2006AJ....132.2360H}.  
They have been shown to have a high rate of hosting exoplanets.
\citet{2007ApJ...670..833J} found that the rate of M stars hosting Jovian planets within 2.5 AU is $1.8\% \pm 1.0\%$ for M dwarfs.
\citet{Dressing2013} claim, with a 95$\%$ confidence lower limit, that the occurrence rate of earth sized planets in the habitable zones of cool stars is 0.04 planets per star.  

In studying the potential habitability of these exoplanets, phenomena such as flaring and the potentially more impactful Coronal Mass Ejection (CME) have to be taken into consideration.  
A CME impacting a planetary atmosphere can cause space weather and other deleterious affects \citep{Lammer2007,Khodachenko2007}. 
Due to tidal locking, the rotation rate of exoplanets residing in the habitable zone of an M dwarf 
is expected to be low \citep{Kite2011}.
The weak magnetic moment of these exoplanets produce a weakened magnetosphere because of the low rotation rate \citep{Khodachenko2007}.
Therefore they are not expected to be as capable at deflecting CMEs as the Earth \citep{Lammer2007}.  
CMEs have the potential to erode the atmosphere when the pressure exerted by the CME compresses the magnetosphere into the atmosphere \citep{Khodachenko2007}.

Solar scaling relations have been used for these phenomena in order to relate stellar flares that are observed to the properties of CMEs which might be occurring with the flares \citep{2012ApJ...760....9A,2013ApJ...764..170D,Osten2015}.  
A few observed stellar counterparts to solar flare phenomena include 
white light flares \citep{1995ApJ...453..464H}, 
nonthermal gyrosynchrotron emission \citep{2006ApJ...637.1016O}, 
and coherent radio emission \citep{2006ApJ...637.1016O}. 
Although the details are still unclear, this multi-wavelength nature of the flare process suggests that both solar and stellar flares share the same basic physical processes. 
This motivates a connection between stellar flares and stellar CMEs in the same way as between solar flares and solar CMEs.

Although applying solar scaling relations provides an easy approach for modeling, there is reason to be skeptical. 
Detailed comparisons of solar and M dwarf flares show a difference in the nature of accelerated particles. 
M dwarf flares produce more accelerated particles than solar flares.
Due to a ``quiescent" nonthermal microwave component during times when no obvious flares are occurring and a perpetuated hot plasma, M dwarf flares produce more accelerated particles.
This suggests that the flare like mechanism of particle acceleration and coronal heating operate at all times in the corona \citep{1996ApJ...471.1002G}. 
Also, solar particles do not penetrate the photosphere deeply enough to reproduce the observed magnitude of the white light stellar flare signal \citep{Kowalski2015} causing difficulties in the application of the standard solar flare model to M dwarfs \citep{2006ApJ...644..484A}.

Despite the fact that stellar flares are routinely observed on different types of stars, clear signatures of stellar CMEs have been less forthcoming. 
In the past, observations of stellar radio bursts at meter wavelengths were limited to those bursts with intensities greater than several hundred mJy \citep{1990SoPh..130..265B}. 
Only a few attempts have been made \citep{1990SoPh..130..391J,1988ApJ...334.1001K,GudelBenz1989,Davis1978} to search for radio emission.
The emission from flare stars at these wavelengths were often assumed to be analogous to the solar type II and type IV radio bursts.
\citet{1990SoPh..130..391J} used the now closed Clark Lake radio telescope to search for radio emission from flare stars in the 30.9 to 110.6 MHz range.
No events were detected in the 30.9 to 110.6 MHz range within limits of 1 Jy (for 1 hr integration) to 50 Jy (for 1s integration).
\citet{1988ApJ...334.1001K} used the 333 MHz capability at the VLA to detect highly polarized emission from YZ CMi.
UV Cet was observed at 333 MHz by \citet{1988ApJ...334.1001K}  and \citet{GudelBenz1989}, but had no detections. 
\citet{Boiko2012} used the ground-based radio telescope UTR-2 to observe flare stars AD Leo and EV Lac.
They claimed, with high confidence, that they found radio emission originating from AD Leo that had burst structure similar to solar type III bursts in the decameter range. 
A stellar analogue of the solar type II burst has not been detected \citep{Leitzinger2011}. 

Traditional solar observations of CMEs use coronagraphs to observe Thomson scattering of photospheric photons off coronal electrons \citep{2006ApJ...642.1216V}.  
This white light emission is faint when compared to the integrated solar disk emission necessitating the use of a coronograph.   
Current astronomical coronagraphs cannot achieve sufficient star contrast to detect a CME at a distance of 1 - 2 R$_\ast$ \citep{Mawet2012} making it unfeasible for CME observations.  
Therefore, only integrated stellar disk emission can be used for observations. 
Flare observations are not affected by this constraint, but the ability to use Thomson scattering to determine CME properties such as its size, mass, speed, and occurrence rate has been removed. 

The type II burst, non-thermal radio emission originating from fast mode magnetohydrodynamic (MHD) shocks,
is an alternative diagnostic \citep{2006GMS...165..207G}.  
The shocks accelerate non-thermal electrons, which in turn produce radio emission at the fundamental and harmonic of the local plasma frequency via well-known plasma processes \citep{2006GMS...165..207G}. 
Type II bursts appear as a slowly drifting radio burst following an exponential path through frequency with time related to the speed of the shock and the density scale height in the corona \citep{2006GMS...165..207G}. 

Typical solar emission frequency drift rates of type II bursts have a range based on starting frequency, shock speed, and characteristics of the corona.  
Starting frequencies of 100, 10, and 1 MHz, result in frequency drift rates of $\sim 10^{-1}, 10^{-3}$, and $10^{-5}$ MHz/s respectively \citep{2006GMS...165..207G}.
The duration in which the type II burst drift from their starting frequency ($\sim100$MHz) down to frequencies below the ionospheric cutoff is many minutes \citep{2006GMS...165..207G}.

Typical solar CME speeds range from a few 100 km/s to about 3000 km/s \citep{2011SoPh..268..195A}.
The probability of having an associated type II burst increases with CME speed.
The association rate reaches 100\% at speeds near 2000 km/s \citep{2008AnGeo..26.3033G}.
Although not all CMEs produce type II bursts, a type II burst is a direct indication that a CME has occurred \citep{2006GMS...165..207G}.
Thus, the type II rate can be use as a lower limit on the CME rate.

Given the coronal characteristics of M dwarfs, these shocks are expected to be observable in the frequency ranges between Earth's atmospheric cut off of $\sim$10 MHz to $\sim$1 GHz.
The $\sim$1 GHz limit corresponds to emission from the base of the stellar corona where electron densities of $\sim10^{10}$ cm$^{-3}$ occur \citep{2004A&A...427..667N}.  
LOFAR is a new facility that is capable of exploring part of this frequency space for detections of transient mass loss.
LOFAR is capable of exploring the frequencies between 10 - 190 MHz.  
LOFAR's millisecond to microsecond time resolution \citep{2013A&A...556A...2V} is a valuable asset for looking for variable-timescale transient objects such as a type II burst.

This paper presents the search for bursting low frequency stellar radio emission and a methodology for interpreting radio observations in concert with flare recordings. 
In Section 2 properties of the star YZ CMi are discussed and the details of LOFAR's beam-formed mode are presented. 
The methodology for data reduction is shared in Section 3.
The results of the observations are discussed in Section 4.1.  
Physical properties derived from the results are examined in Section 4.2.
Section 4.3 presents the methodology for estimating frequency drift rates of observed type II bursts from the physical properties found in Section 4.2.  
An examination of the assumptions used in analysis is shared in Section 4.3.
A methodology for connecting flares and CMEs to further improve analysis for future experimentation is proposed in Section 5.
Section 6 concludes.

\section{Observations}

\begin{deluxetable}{ccccccccc}
	\rotate{}
	\tabletypesize{\footnotesize}
	\tablecaption{LOFAR Observations and Information}
	\tablecolumns{9}
	\tablewidth{0pt}
	\tablehead{
		\colhead{Observation} & \colhead{Antenna} & \colhead{Observed} & \colhead{Width per}  & 	\colhead{Start Time} & \colhead{$\#$ of (Inner)} &\colhead{Total Time} & \colhead{Sampling} & \colhead{Number of}\\    
		\colhead{ID} & \colhead{Set} & \colhead{Frequencies [MHz]} & \colhead{channel [KHz]} &\colhead{[UT]}  &\colhead{Core Stations} &\colhead{Observed [s]}  &\colhead{Time [s]} &\colhead{Channels} 
	}
	\startdata
		98796 & LBA & 29.980-77.636 & 0.763 & 2013-02-25 03:45:00 & 24 & 10799.81 &0.131072 & 62464\\
		98950 & LBA & 10.448-58.105 & 0.763 & 2013-02-26 03:45:00 & 23 & 10799.81 & 0.131072& 62464\\
		99198 & HBA & 114.939-162.596 & 3.052 & 2013-02-28 03:45:00 & 2 $\times$ 21  & 10779.91 & 0.131072 & 15616\\
		184329 & HBA & 114.939-162.596 & 3.052 & 2013-11-01 03:45:00 & 2 $\times$ 20 & 10779.91 & 0.131072 & 15616\\
		184330 & HBA & 114.939-162.596 & 3.052 & 2013-10-01 03:45:00 & 2 $\times$ 20 & 10779.91 & 0.131072 & 15616 \\
	\enddata
	\tablecomments{Comprehensive table of various aspects of the observations used for the analysis in this paper.  The HBA filter is between 110-190 MHz, and the LBA filter is between 10-90 MHz.
	The pointing is for the target YZ CMi. 
	For the off beam, HBA observations have an offset of 10 degrees in Declination while LBA observations have an offset of 15 degrees in Declination. 
	LBA observations were taking with the LBAOUTER layout.
	HBA observations used the HBAdual configuration.
	  }
\end{deluxetable}

The primary observing target was YZ CMI, an active M dwarf, because of its high optical flare rate and previous flare observations allowing for detailed knowledge of its stellar parameters.  
It has a radius of $\sim$0.36 R$_{\odot}$ \citep{1992ApJ...397..225M}, mass of $\sim$0.34M$_{\odot}$ \citep{1987PASAu...7..197L}, and is at a distance of 5.93 pc \citep{1997A&A...323L..49P} from Earth.  
The magnetic and plasma properties have been determined from previous observations: the magnetic field strength at flare sites between 50-100G \citep{2007MNRAS.379.1075R}, a coronal temperature of $10^{7}$ K \citep{1993ApJ...418L..41D}, and a best value of quiescent state coronal density of $\sim3\times10^{10}$ cm$^{-3}$ \citep{2004A&A...427..667N}.
YZ CMi has an optical flare rate of 0.4 flares/hr for flares above U-band energy of 5$\times$10$^{31}$ ergs (U-band is centered around 364 nm) \citep{Lacy1976}. 
Also, previous observations of YZ CMi have detected low frequency (1500 and 300 MHz) microwave events \citep{1988ApJ...334.1001K}.  

By observing YZ CMi at radio frequencies, the presence of CMEs can be confirmed and their properties may be constrained via detections of type II radio bursts.  
LOFAR's tied-array Beam Formed Mode was used for the pilot project LC0\_013 for this purpose.  
Instead of producing interferometric visibilities, LOFAR's beam-formed modes can combine the LOFAR collecting area into ``array beams", either the coherent or incoherent sum of multiple station beams \citep{Stappers2011}.
These data are used to produce time-series and dynamic spectra for high-time-resolution studies (intensities as a function of time and frequency). 
The Coherent Stokes sub-mode produces a coherent sum of multiple stations ("tied-array" beam) by correcting for the geometric and instrumental time and phase delays \citep{2013A&A...556A...2V}.  
This produces a beam with restricted field-of-view ($\sim$5 deg maximum), but with the full, cumulative sensitivity of the combined stations \citep{2013A&A...556A...2V}.  
These observations used LOFAR's core observation stations and did not utilize any of the remote or international stations.  
Between 20 and 24 core stations were coherently combined using LOFAR's Blue Gene/P correlator/Beam-former to form a tied-array beam for high time and frequency resolution observations.  

High Band Antennas (HBA) with frequency ranges between 110-190 MHz or Low Band Antennas (LBA) with frequency ranges between 10-90 MHz were used for observations. 
LBA observations used the LBAOUTER configuration which is best used for observations below 40 MHz. 
HBA stations in the core are arranged in 2 $\times$ 24-element fields.  
The two fields can be used either as individual stations (HBAdual) or their signals can be combined to form one 24 element field \citep{2013A&A...556A...2V}. 
The HBAdual configuration was used for these observations.

The five observations presented in this paper took place on February, October, and November of 2013 and started at 03:45:00 UT on their respective days.  
The 15 hours of observations were split between either HBA or LBA modes in three hour intervals as listed in Table 1. 

Observations include simultaneous on-beam (science target) and off-beam (an adjacent blank patch of sky) signals.  
Off beams had a declination offset of 10 degrees for HBA and 15 degrees for LBA relative to the on-beam.
The frequency channels were averaged in frequency to create 244 subbands. 
HBA observations were averaged over 64 channels while LBA were averaged over 256 channels to create a subband.
This creates the same effective bandwidth per subband of 195 kHz for both HBA and LBA observations.   
See Table 1 for a comprehensive summary of the observations.

The sensitivity of a combined set of coherent tied-array beams is equivalent to that obtained from the sum of the collecting areas of all the stations being combined \citep{Stappers2011}.
By using the sensitivity of N equal stations and the System Equivalent Flux Densities ($\Delta S$) listed on LOFAR's sensitivity web page
\footnote{\label{note1}https://www.astron.nl/radio-observatory/astronomers/lofar-imaging-capabilities-sensitivity/sensitivity-lofar-array/sensiti}
 with a integration time ($\delta t$) and bandwidth ($\delta \nu$), a theoretical sensitivity can be determined $\left(\Delta S = \frac{S_{sys}}{\sqrt{N(N-1)2\delta t \delta \nu}}\right)$.
These observations did not include a calibration target for absolute calibration so theoretical limits will be imposed as upper bounds for sensitivity. 
Table 2 lists the approximate sensitivities, as determined by the LOFAR sensitivity web page and number of antennas presented in Table 1.  
The values in Table 2 are approximate, and represent an underestimate of the true limits.

\begin{deluxetable}{cccccccccc}
	\tabletypesize{\footnotesize}
	\tablecaption{LOFAR Sensitivity}
	\tablecolumns{3}
	\tablewidth{0pt}
	\tablehead{
		\colhead{Observation} & \colhead{Antenna} & \colhead{Sensitivity\tablenotemark{1}}\\        
		\colhead{ID} & \colhead{Set}  & \colhead{[mJy]} }
	\startdata
		98796 & LBAOUTER &  327 - 474\\
		98950 & LBAOUTER  &  449 - 251\\
		99198 & HBAdual & 12 - 10  \\
		184329 & HBAdual &  13 - 10 \\
		184330 & HBAdual &  13 - 10 \\
	\enddata
	\tablecomments{The bandwidth used for calculations is 195 kHz and the integration time was 60 second.  
}
\tablenotetext{1}{
The order of the sensitivities listed matches the frequencies listed in Table 1 for each Observation ID.
}
\end{deluxetable}

\section{Data Reduction}

In order to reduce the data, an average of frequency channels for both LBA and HBA are taken.  
Frequency is averaged, as explained above, and the time scale is averaged from the 0.131s integration time used in observations to $\sim$1s timescales or longer.
To account for any frequency dependent signal intensity variations, each frequency subband (244 subbands in total) is scaled by the median value of that subband providing a scaled unit-less intensity.  


\begin{figure}
         \includegraphics[scale=0.45, clip=true, trim=1.5cm 6.5cm 2cm 7cm]{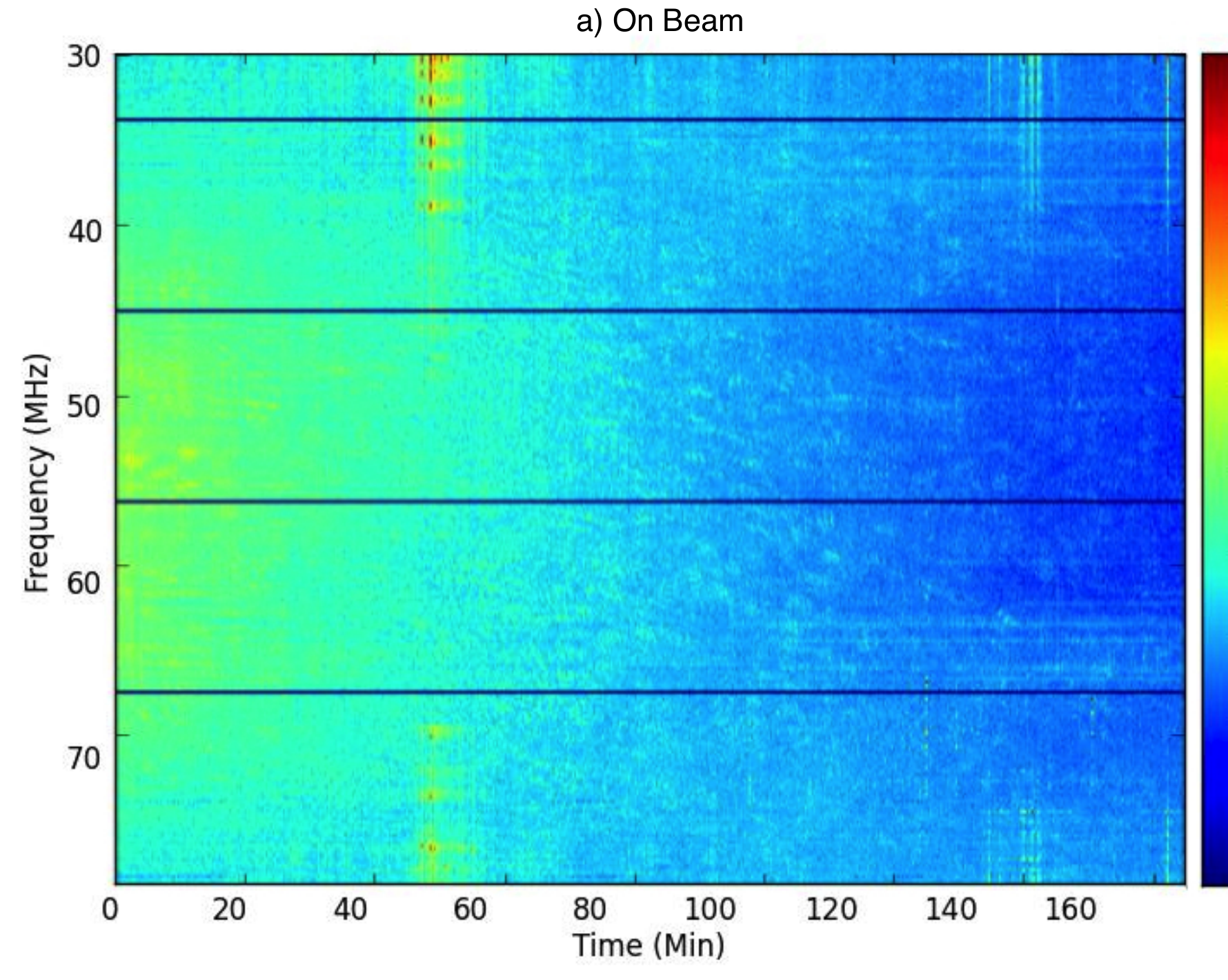} 
         \includegraphics[scale=0.45, clip=true, trim=1.5cm 6.5cm 1.5cm 7cm]{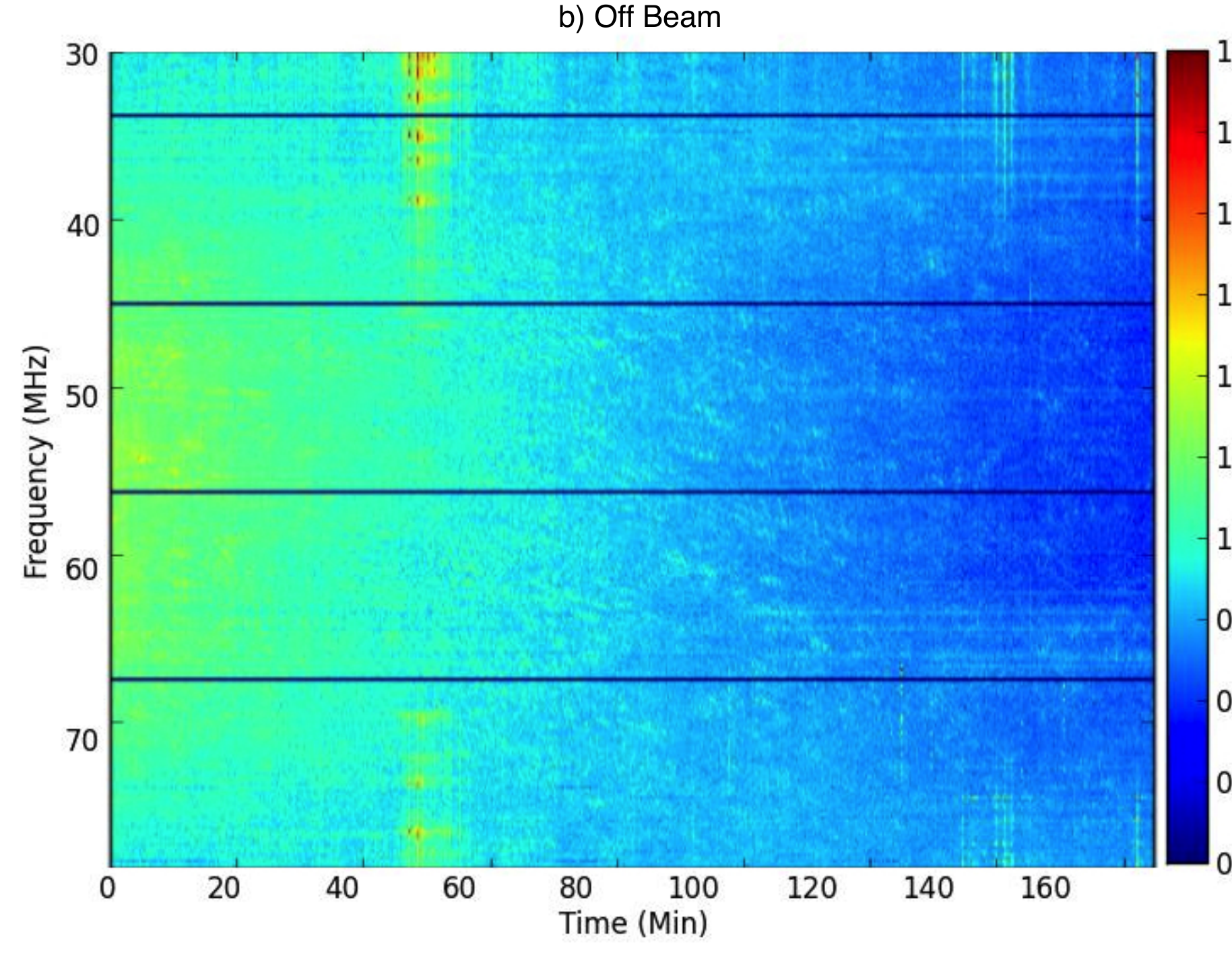}
         \caption{98796 LBA beam comparison to display RFI and potential signal detection.  Panel a) shows the ``on beam" while panel b) is the ``off beam". 
        Each frequency subband is scaled by the median value of that subband providing a scaled dimensionless intensity.  
        }
        \label{fig:98796Beams}
\end{figure}

Figure \ref{fig:98796Beams} shows the data after initial averaging in time and frequency.  
 A variety of transient and pseudo-persistent effects plague the observed spectra. 
One such aspect, not displayed here, is the ionospheric frequency cutoff ($\sim$10 MHz) affecting the lowest observable frequencies of LBA spectra.
Any prominent characteristic present in both beams is a sign that the signal source is inherently terrestrial as it is independent of pointing and is considered to be Radio Frequency Interference (RFI).  
Therefore any shared characteristics between the on-beam and off-beam shall be flagged before proceeding to the next step (and after averaging has occurred).
  
\begin{figure}
\centering
\includegraphics[scale=0.8,clip=true]{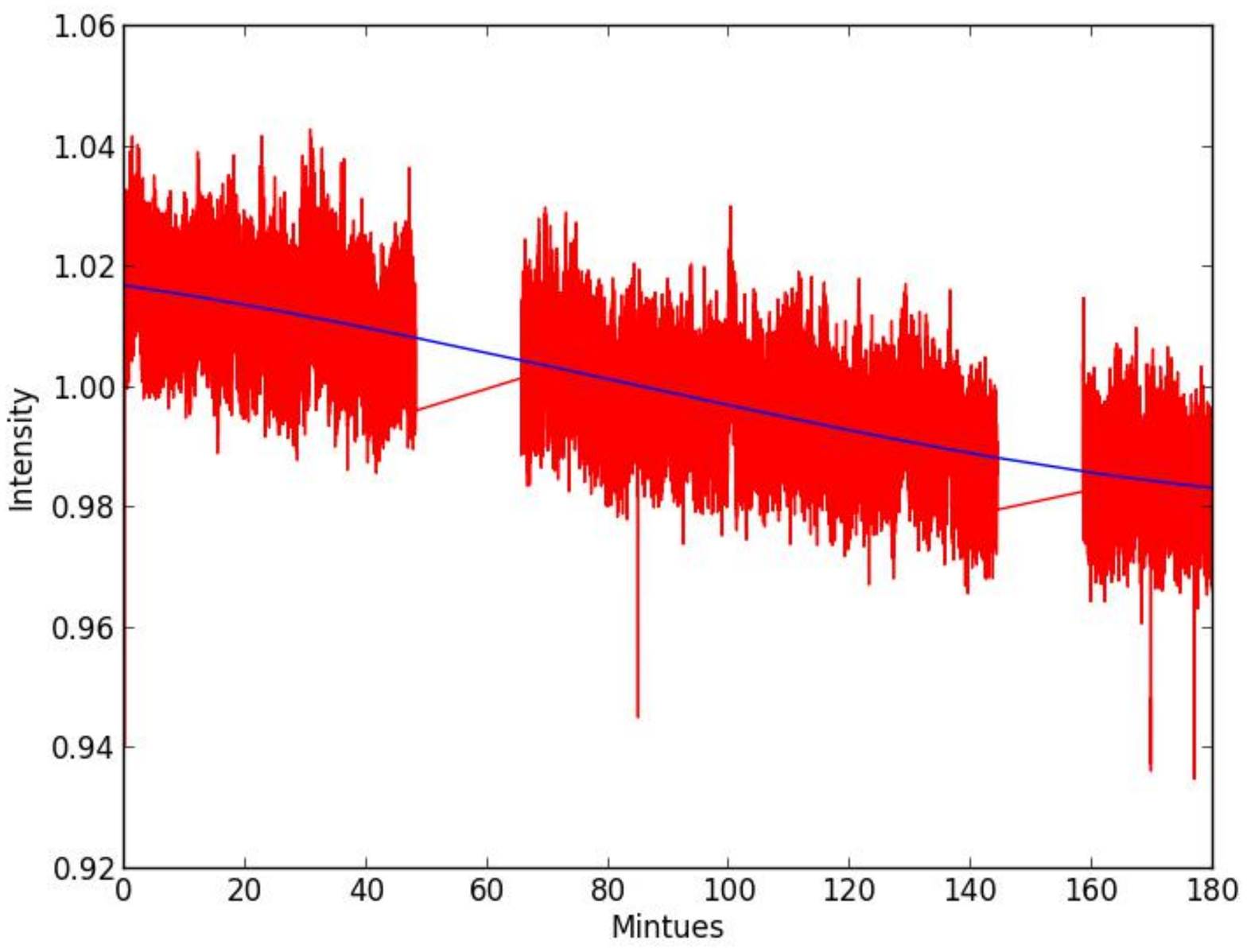}
\vspace{-6cm}
\caption{LBA data set 98796 is used to display an example of creating a line of best fit for determining inherent frequency-dependant noise levels.
The displayed frequency is for $\nu = 74.90$ MHz. 
Gaps occur from prior RFI flagging.}
\label{fig:230avt}
\end{figure}

\begin{figure}
\centering
\includegraphics[scale=0.5,clip=true]{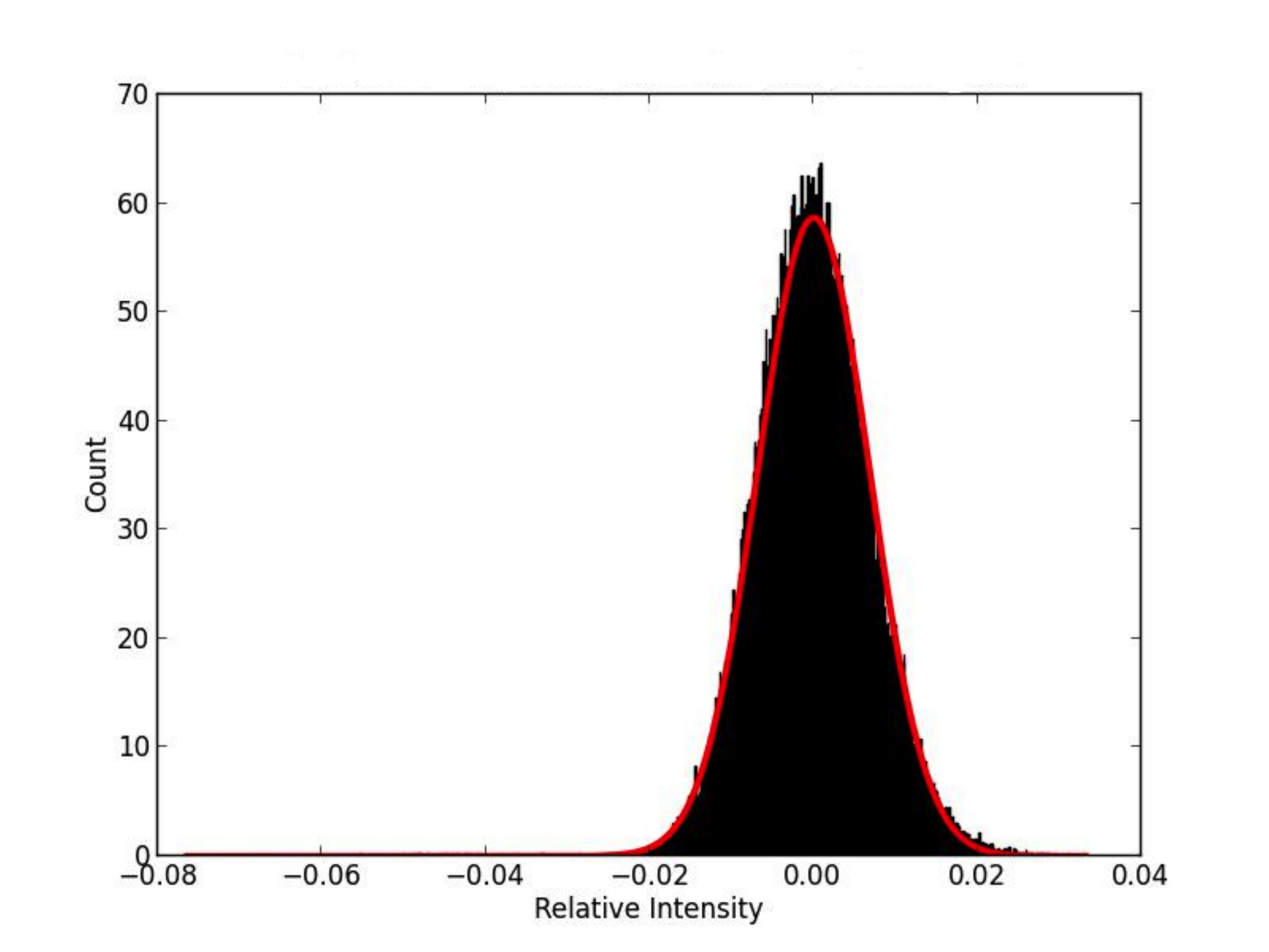}
\caption{
A histogram of Figure \ref{fig:230avt}'s deviations from the line of best fit is displayed with its modeled Gaussian fit.
The Gaussian is used in order to obtain a noise level for this frequency channel ($\nu = 74.90$ MHz).
This particular example has a standard deviation of 0.068 relative intensity.
}   
\label{fig:230NOISE}
\end{figure}

To rigorously identify statistically-significant events throughout the data, a frequency-dependent noise level is determined for the data.
Figure \ref{fig:230avt} shows graphically the start of this process. 
The first step is to model how the sensitivity in each frequency channel is changing in time to help remove systematic variations.  
A 3$^{rd}$ degree polynomial is used to generate the baseline.  
The deviations from this baseline are then calculated and plotted in Figure \ref{fig:230NOISE}.
Using RMS variations from this fit, a Gaussian distribution can be modeled to determine outliers and determine frequency-dependent noise level ($\sigma$).  

\begin{figure}
\centering
\includegraphics[scale=0.8,clip=true, trim=0cm 0cm 0cm 4cm]{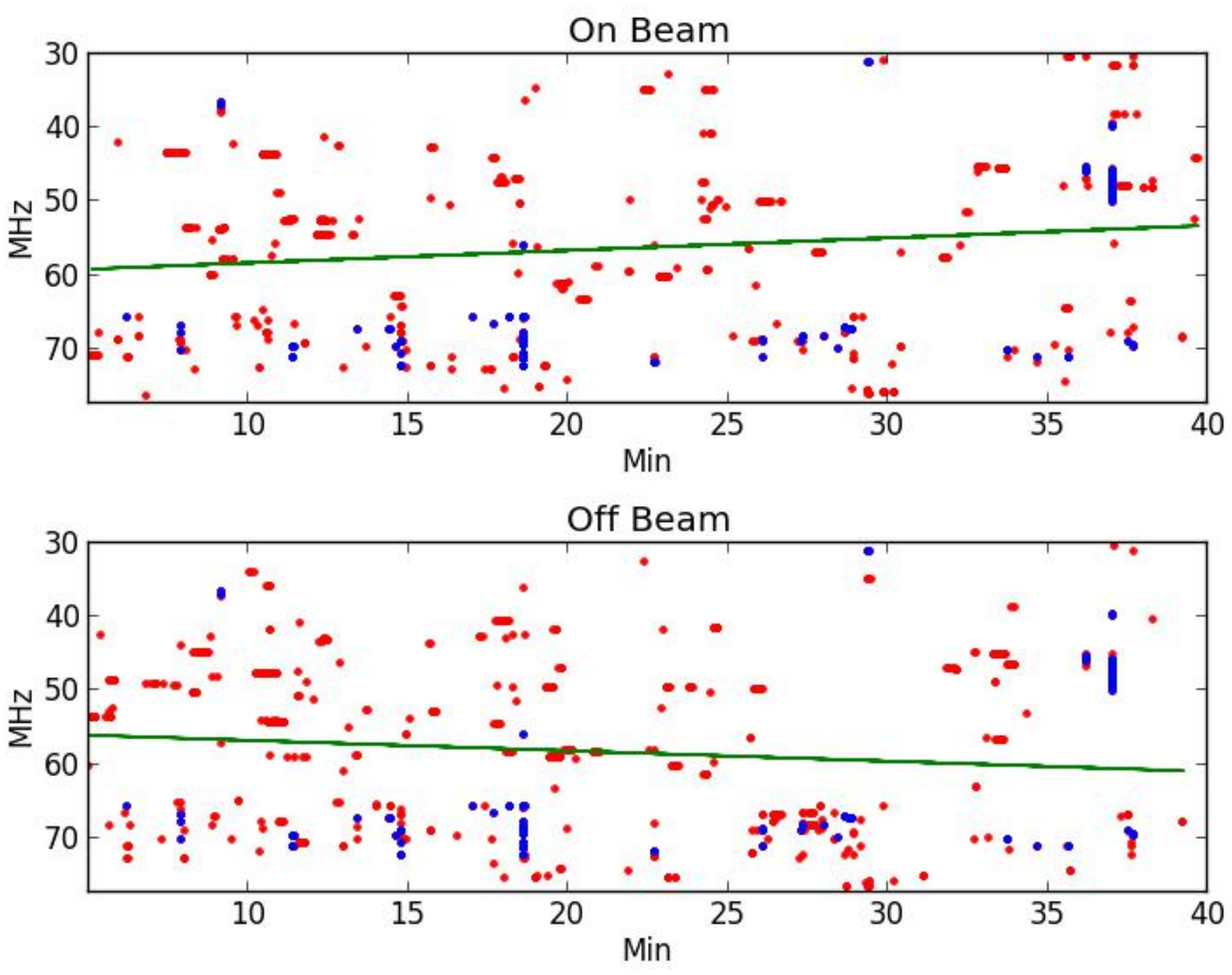}
\vspace{-6.5cm}
\caption{This displays a drift rate calculation for non flagged points above a 5$\sigma$ cut off for both on and off-beams for LBA run 98796.  The two images display the points as red dots with a calculated green slope through it. Blue colored points are points at which both the on and off beam had a statically significant value and should be treated as a potentially terrestrial signal.
       The frequency range is over the entire potential frequency range (30-77.6 MHz) but the time axis only covers about 35 mins of a potential 180 min total range.  The slope is derived from 3545 remaining points that passed a 5$\sigma$ threshold of the original 54931 points in this plotted range.  No points are expected to remain at this threshold if no signal is present.  The on beam in this case display a drift rate of $-2.82\times10^{-3}$ MHz/s while the off beam has a slope of $2.36\times10^{-3}$ MHz/s.}
\label{fig:4plot}
\end{figure}

After repeating for both on and off-beams, the two are then compared to look for potential signals.  
Figure \ref{fig:4plot} shows a comparison of on and off-beams over the entire frequency range for a 35 min portion of the data set.  
The points displayed are only those remaining after a 5$\sigma$ threshold has been enforced in order to find outliers and a linear drift rate slope has been superimposed on top. 
For reasons explained below, the type II burst is expected to have a duration of tens of minutes.
The time range imposed on Figure \ref{fig:4plot} is chosen to be long enough so that the entire duration of an expected type II on YZ CMi would be observed.
This time is also long enough such that if the conditions differ from what is expected, a partial signal should still be present which would suggest the need to use a longer time segment.

\section{Results and Discussion}
\subsection{Results}

These observations have yielded no plausible detections of drifting low-frequency (type II) radio bursts on YZ CMi and thus no indirect CME detections either. 
A type II burst is the result of a CME creating an MHD shockwave as it travels though the stellar atmosphere.  
Since not all CMEs create type II bursts, the frequency of type II's will always be less than or equal to the frequency of CMEs. 
Therefore the frequency of type IIs is a lower bound for the frequency of CMEs.  
Assuming the rate of shocks is a lower limit to the rate at which CMEs occur, no detections in a total of 15 hours of observation places a limit of $\nu_{type II} < 0.0667$ shocks/hr $ \leq \nu_{CME}$ for YZ CMi due to the stochastic nature and of the events and limits of observational sensitivity.  

Section \ref{MethodologyForConnection} proposes a methodology for performing simultaneous observations of flares (optical observations of a star) while observing for type II shocks (radio observations of a star) in the manner presented in this paper. 
The results of this methodology is to advance and improve further experiments by providing a more in-depth list of constraints on CME properties than what can be achieved by radio observations alone.  

\subsection{Limits}
\label{lims}

\citet{2006ApJ...650L.143Y} showed that X-ray solar flare energies above 10$^{31}$ erg in the GOES bandpass (1-8 \AA) have a 100\% correspondence between solar flares and CMEs.  
Adjusting the flare energy limit placed by \citet{2006ApJ...650L.143Y} to be representative of u-band energy \citep{Osten2015}, this would imply the limit would be $\sim2\times10^{31}$ erg.
YZ CMi has an optical flare rate of 0.4 flares/hr for flares above U-band energy of 5$\times$10$^{31}$ \citep{Lacy1976}. 
Assuming that the solar relationship holds for M dwarfs, this would imply that YZ CMi would have an  expected CME rate, while not accounting for observational constraints, of $\nu_{CME} \ge$ 0.4 CMEs/hr. 
This is much greater than the upper limit 0.0667 shocks/hr $ \leq \nu_{CME}$.
An expectation of 1.2 optical flares per 3 hour observation session leads to a predicted 1.2 shocks per 3 hr observation period for YZ CMi. 
The Poisson statistics state that the probability of getting 0 detections is $\sim0.25\%$.
This is suggestive of a more definitive result as opposed to an unfortunate low-activity period for YZ CMi during observations.  

With the lack of confirmed detections of stellar type II bursts, the sensitivity of LOFAR shall be used to determine the limits on ($T_{b}$) and source size ($\Omega$) of the Type II bursts.

The flux density can be estimated from the type II burst brightness temperature and the size of the emitting region.  
The flux density ($S_{\nu}$) is proportional to the brightness temperature ($T_{b}$) and solid angle ($\Omega$) ($S_{\nu} \propto \int T_{b}\nu^{2} d\Omega$).
The brightness temperature T$_{b}$ can be estimated from models of the waves involved \citep{BenzThejappa1988} and is about $10^{14}$K for solar type II bursts. 
Stellar coherent bursts show evidence for larger $T_{b}$ \citep{2006ApJ...637.1016O}; therefore the stellar case is taken to be $T_{b} \geq 10^{14}$ K.

\begin{figure}
\centering
\includegraphics[scale=.8,clip=true, trim=0cm 0cm 0cm 4cm]{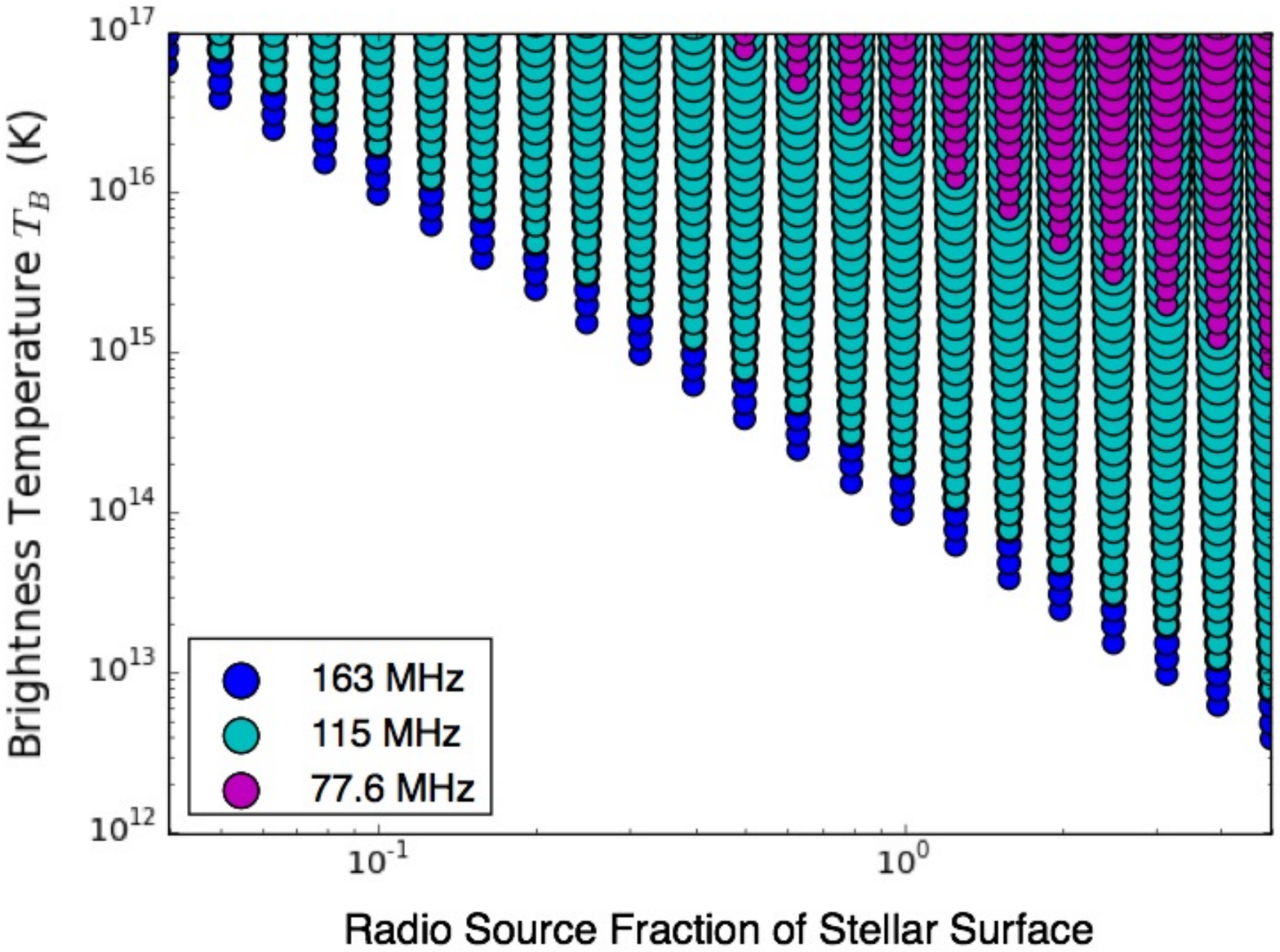}
\vspace{-6cm}
\caption{The detectable levels of radio emission at several frequencies covered by the observations for combinations of brightness temperature and radio source extent. 
The noise parameters in Table 2 are used to determine sensitivities.  
Blue points represent a frequency of 163 MHz and cyan points represent 115 MHz, and purple points represent 77.6 MHz.  
The points presented are ones that have a signal-to-noise (S/N) ratio greater than 5 and the size of the point is scaled logarithmically to the S/N ratio.
The x-axis ranges from values of 0.04 to 5 times the area of the stellar disk.  
}
\label{fig:sens}
\end{figure}

To have a better understanding of where expected Type II bursts will produce detectable results, Figure \ref{fig:sens} displays flux densities that are a factor of 5 above their respective noise levels at LOFAR for various frequencies.
Table 2 is used for the frequency-dependant noise levels used in calculations for Figure \ref{fig:sens}.
In order to have detectable bursts, large angular scale, and high brightness temperature CMEs are required.
Very bright bursts (10$^{16-17} $K) may still be undetectable if they do not fill a large enough fraction of the stellar surface area.   
This is likely a contributor to the lack of detections as even a CME that produced a type II shock in the correct parameter space would have the additional constraint of having high enough brightness temperature and/or large enough angular size to be detected by LOFAR.  

\citet{Susino2015} studied a GOES M2.6 class solar flare (medium-large flare) and found that in the early phases (2 - 4 R$_{\odot}$) the whole shock surface is completely super-Alfv\'{e}nic, while later on it becomes super-Alfv\'{e}nic only at the nose.  
\citet{Byrne2012} describes the angular spread of a CME as a function of the radial height (r) above the Sun with a power-law expansion $(\Delta\theta(r) = \Delta\theta_{0} r^{0.22}$, where $ \Delta\theta_{0} = 26^{\circ}$) .  
The estimated angular spread of a fully super-Alfv\'{e}nic CME at 4$R_{\odot}$ is $\sim35^{\circ}$.
The angular scale of the transient shock at this height (4R$_{\odot}$) would be $\sim$1.6 times the size of the solar disc.  
Assuming that this relation holds true for M Dwarfs as well, it can be expected that a CME would have an area 1.6 times the stellar surface.  
Assuming a head on CME (where the entirety of the CME is visible), Figure \ref{fig:sens} shows that at roughly 1.6 times the stellar surface area, the minimum brightness temperature of $\sim 5 \times 10^{13}$ K for higher would be required for LOFAR's observable frequencies. 
This is near the $T_{b} \geq 10^{14}$ K limit set previously for expected brightness temperatures.   

\subsection{Estimation of Drift Rates}

The characteristic shape of a Type II burst is a consequence of the emission mechanism for this event. 
Langmuir waves are produced by electrons which are accelerated by the MHD shock generated from the movement of the CME through the corona and radiate via the local plasma frequency and its harmonics.
As the shock moves through the corona, the frequency of the accompanying radiation also varies.
The frequency will vary as:
\begin{equation}
\frac{d\nu}{dt}=\frac{\partial\nu}{\partial n}\frac{\partial n}{\partial h}\frac{\partial h}{\partial s}\frac{\partial s}{\partial t} 
\end{equation}
where $\nu$ is frequency, $n$ is electron density, $h$ is radial height above the star, $s$ is distance along the path which the shock travels, and $t$ is time.  
The drift rate is thus composed of four terms describing the changing environment around the shock.  
As the shock propagates away or towards the star, the local electron density $(n)$ will change.
The shock will emit via the local plasma frequency $(\nu_{p} = \sqrt{\frac{ne^{2}}{\epsilon_{0}m_{e} } } $ in SI) where $n$ is electron density, $m_{e}$ is the electron mass, e is the electron charge, and $\epsilon_{0}$ is the permittivity of free space. 
An outward moving shock will emit at a lower frequency as it travels towards lower densities $\left(\nu_{p} \propto \sqrt{n} , \frac{\partial \nu}{\partial n} = \frac{\nu}{2n} \right)$. 
The second term describes how the density changes as a function of height radially above the stellar corona. 
The start frequency gives information on height in the stellar atmosphere via a barometric model $\left(\frac{\partial n}{\partial h} = \frac{- n}{H_{0}} \right)$. 
The final two terms describe the path and speed the CME takes as it travels. 
The CME does not necessarily travel perpendicularly outward and so $\frac{\partial h}{\partial s} = cos\theta$ describes how the vertical height changes as a function of the path traveled.
Here $\theta$ is the angle at which the shock is traveling relative to the radial direction.
The CME will travel a sufficiently small distance over the course of our observations such that we may assume that it will have a constant speed v$_{s}$.  
Taking this all into consideration leads to a simplified expression for the drift rate of:
\begin{equation}\label{eq:nudot}
\frac{d\nu}{dt} = \left(\frac{\nu}{2n}\right)\left(-\frac{n}{H_{0}}\right)\left(cos\theta\right)\left(v_{s}\right) = -\frac{\nu v_{s} cos\theta} {2H_{0}}  
\end{equation}
where $\nu$ is frequency, $v_s$ is velocity of the shock, $\theta$ is the angle at which the shock path is oriented relative to the radial direction, $H_{0}$ is the density scale height.
 
The emitted frequency from the type II burst is an observable and therefore can be directly measured. 
Measuring how the observed frequency changes with time provides a direct measurement for the drift rate.
Density scale height can be estimated from coronal base temperatures, leaving v$_{s}cos\theta$ as the remaining unknown parameter.
Assuming a perpendicular path $\left( cos\theta \approx 1\right)$, a detection of a drifting radio burst provides a constraint on the minimum speed of the CME shock and therefore the CME itself.   

\begin{figure}
\centering
\includegraphics[scale=0.7,clip=true, trim=0cm 0cm 0cm 4cm]{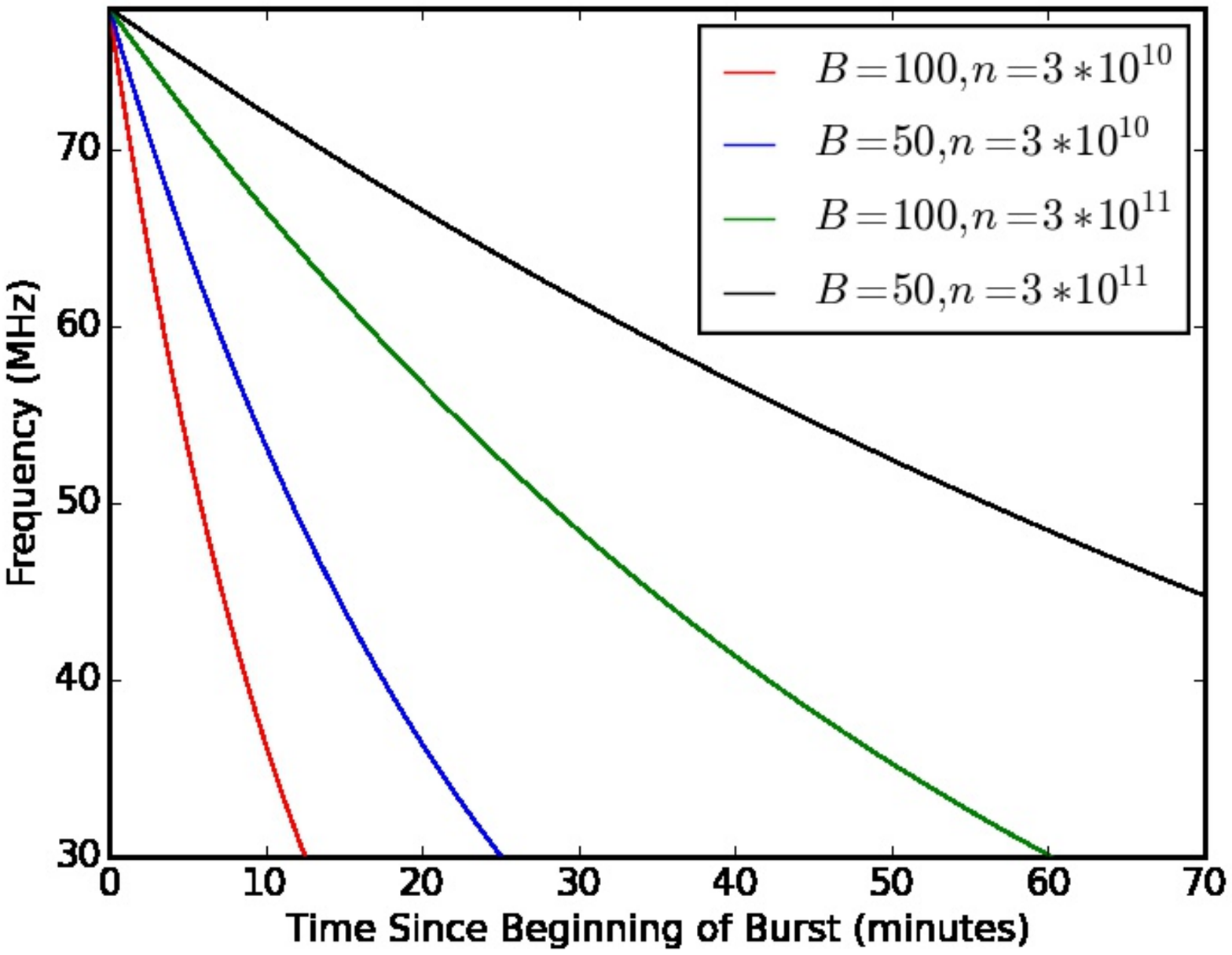}
\vspace{-4cm}
\caption{
The frequency-time signal expected for a Type II burst under various parameters.
Only the intrinsic motion through the outer stellar atmosphere is plotted as ISM signal dispersion effects are not taken into account as they are negligible.  
The shocks are calculated for the LBA window between 78 to 30 MHz akin to observation 98796.  
The shock is formed at a height required to have an initial frequency of 78 MHz (initial shock height will be different for each set of initial conditions) and thus the path described will be the slowest potential path through the frequency range possible.
} 
\label{fig:signal}
\end{figure}

In order to evaluate drift rates, the known stellar parameters of YZ CMi stated above will be used for calculations. 
The parameters are: coronal density scale height will be $1.67 \times 10^{3} T$ cm with $10^{7}$K \citep{1993ApJ...418L..41D}, radius R$_{\ast}=$0.36$R_{\odot}$ \citep{1992ApJ...397..225M}, mass $0.35M_{\odot}$, magnetic field strength at base of the corona $B_{0}=100$ G \citep{2007MNRAS.379.1075R}, and $n_{0}=3\times10^{10}$ cm$^{-3}$  \citep{2004A&A...427..667N} for number density. 
The magnetic scale height H$_{B}$ for an exponential field is assumed to be comparable to R$_{\odot}$ for the low-order magnetic field.  
The Alfv\'{e}n  $\left(v_{A} =  2.03 \times 10^{11} \frac{B}{\sqrt{n}} \right)$ speed is used as the minimum speed required for a shock to be generated. 
An Alfv\'{e}n speed scale height  $\left(v_{a} \propto v_{0}e^{R/R_{0}} \right)$ can be estimated using an effective scale height equal to $\frac{2H_{n}H_{b}}{2H_{n}-H_{b}} = 9.2\times10^{10}$cm.  
Taking account of all stated values, Table 3 displays various values of drift rates and initial emission frequencies for various shock formation distances.  

Figure \ref{fig:signal} shows Type II bursts generated in various initial stellar conditions (densities and magnetic field strengths at base of corona) that could generate a signal in the frequency range for the LBA observation (30-78 MHz) using plausible coronal conditions for YZ CMi. 
The bursts shown are formed such that their start frequency is 78 MHz.
Each shock will form at the same local density but the radial height that will correspond to this density will depend on the initial base coronal density.
Since this shock is required to be super-Alfv\'{e}nic, the magnetic field (B) will be used to determine the minimum speed the radio burst can travel.  
The red and the blue lines on this figure are also used as a rough guide for what the true shape of a type II burst is expected to be for YZ CMi.  
This is intended to show the longest possible duration of a type II burst under various stellar conditions. 
The duration of the type II burst can span from $\sim$12 mins to over 100 mins based on the stellar conditions alone.
A measured frequency, a frequency drift rate determined from observations,  and prior knowledge about the coronal temperature to determine a barometric density scale height can be used with equation 2 to extrapolate the velocity of the shock.

\begin{figure}
\centering
\includegraphics[scale=0.7,clip=true, trim=0cm 0cm 0cm 4cm]{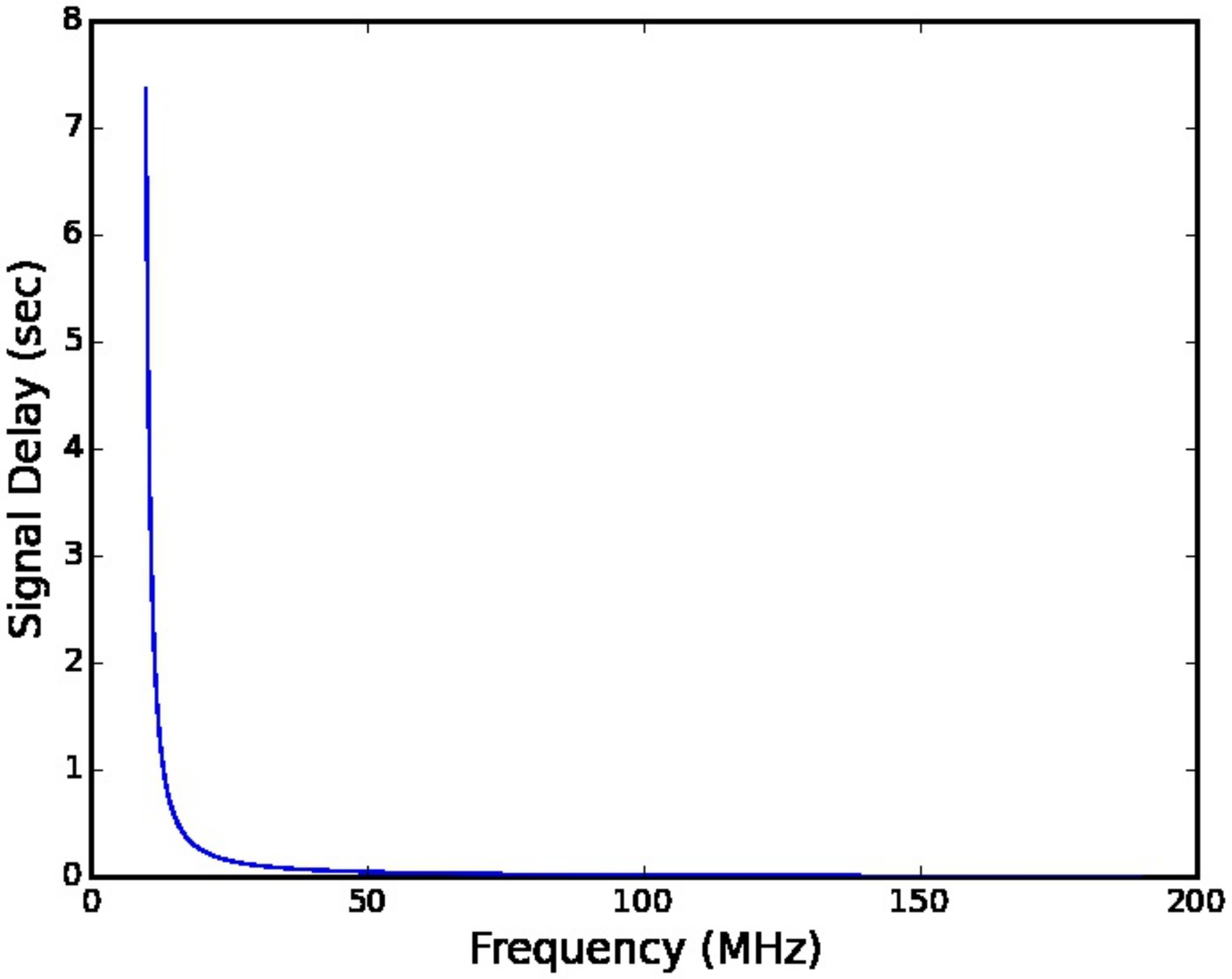}
\vspace{-4.5cm}
\caption{The signal delay as a function of frequency.  The delay time is measured relative to the arrival time of 190 MHz. This curve assumes a starting frequency of 190 MHz for an M dwarf with density at the base of the corona of $n_{0}=10^{10}$ cm$^{-3}$ and magnetic field $B_{0}=100$ G. 
}
\label{fig:delay}
\end{figure}

One final factor to be mindful of is the effects of signal dispersion through the Interstellar Medium (ISM).  
As the signal is propagating through the ISM plasma, a frequency-dependent time delay is introduced. 
The delay between any two frequencies is given as:
\begin{equation}
\Delta\tau = \frac{e^{2}}{2\pi cm_{e}}\left[ \frac{1}{\nu_{1}^{2}} - \frac{1}{\nu_{2}^{2}}\right] \int_0^L N(\ell) d\ell
\end{equation}
 where $\Delta\tau$ is the time delay between the two frequencies $\nu_{1}$ and $\nu_{2}$, the ISM particle density is $N(\ell) \approx0.03$ cm$^{-3}$ \citep{Shu1991}, the distance L to YZ Cmi is 5.93 pc, c is the speed of light, and the electron charge and mass are $e$ and $m_{e}$ respectively.   
The HBA observations (115 - 163 MHz) will have maximal delays of 0.029 seconds, which is insignificant for this analysis.  
The LBA runs 98796 (30 - 78 MHz) and 98950 (10-58 MHz) will have maximal delays of 0.7 and 6.7 seconds respectively.
Any detections in the lowest frequency observations would require that this effect be accounted for to provide accurate drift rate measurements, however it is a small correction.  
As it is difficult to see its impact directly, Figure \ref{fig:delay} shows the signal delay as a function of frequency to emphasize the impact of the delay on lower frequencies.  
This curve assumes a starting frequency of 190 MHz for an M dwarf with density at the base of the corona $n_{0}=10^{10}$ cm$^{-3}$ and magnetic field $B_{0}=100$ G. 

\subsection{Examination of Assumptions}

There are three primary assumptions made for this experiment.  
The first is that the increase in flare rate of an M dwarf relative to the Sun is partnered with an equal increase in rate of CME occurrence. 
Simply, this is the assumption that solar scaling relations between the Sun and M dwarfs hold.  
Possibilities such as an overlying magnetic field structure which is preventing the ejection itself, as has been observed for large solar flares lacking a respective CME \citep{Sun2015}, or perhaps the existence of an alternative mode for energy release assuming the role of a CME are not considered.  

The second assumption is that a shock will form, given sufficient conditions, as the result of a CME passing through the corona.  
If overlying fields are preventing the CME from ever reaching super-Alfv\'{e}nic speeds, or if the densities are being changed in some unknown manner, then no shock will ever form to result in the type II burst required for detection.  

Finally, it is assumed that when a CME has occurred that has created a type II burst, that the burst is detectible. 
As discussed in section \ref{lims}, the signal is dependent on the frequency, brightness temperature $T_{b}$, and the angular size $\Omega$.  
Conditions could be such that any CMEs released are only generating MHD shocks further away from the base of the stellar corona resulting in type II bursts at frequencies below the magnetospheric cutoff and would be undetectable using ground based observatories.  
Another reasonable alternative is that the brightness temperature and/or the angular size are too small to produce a detectable signal.  
It may be the case that the shocks are all very bright but so angularly small that they still are undetectable, or the reverse case that they cover quite a large area but are too cold to be seen.

\section{Method for Connecting Flares and CMEs}
\label{MethodologyForConnection}

On the Sun, flares and CMEs are closely related phenomena; they both occur as the result of magnetic reconnection and manifest together in eruptive events.  
Magnetic reconnection occurs when highly stressed and opposing magnetic field lines are pressed together.  
The field lines break and reconnect into a new configuration allowing the release of energy and/or mass.    
\citet{2006ApJ...650L.143Y} showed that X-ray solar flare energies above 10$^{31}$ erg in the GOES bandpass (1-8 \AA) have a 100\% correspondence between solar flares and CMEs.  
Flares occurring on M dwarfs typically have energies between $10^{28}$ to $10^{34}$ erg and have mean flare rates that are 10$^{4}$ - 10$^{5}$ times higher than on the Sun \citep{Lacy1976,2010ApJ...714L..98K}.
Assuming the relationship holds for M dwarfs, this would imply a large number of M dwarf flares would have the 100\% correspondence of flares to CMEs.

What type of characteristics will this associated CME have for a flare of a certain energy?
Physically driven arguments show that in the solar case there is a rough equipartition between the total radiated energy of the flare and the mechanical energy of its associated CME
\citep{Emslie2005,Osten2015}.  
\begin{equation}
\label{eq:equipar}
\frac{1}{2} M_{CME} v^{2} = \frac{E_{rad}}{\epsilon f} 
\end{equation}
Where $f$ is the fraction of the bolometric flare emission appropriate for the waveband in which the energy of the flare is being measured.
In this case, $f=f_{u} = 0.11$ for a u-band (300-430 nm) optical observation \citep{Osten2015}, $\epsilon$ is a constant of proportionality	 $\approx 0.3$, $v$ is the velocity of the CME, $M_{CME}$ is the total mass of the CME, and $E_{rad}$ is the radiated optical flare energy in the u-band.

Given a measured flare energy, and this relation, there still exists a mass-velocity degeneracy for a CME occurring with a flare.  
However, radio measurements of the CME provide a velocity constraint which can be used to break the degeneracy.  
Assuming this solar relation holds for the stellar case, a method for constraining CME mass, that was previously impossible using radio observations alone, is obtained.  

An experimentally found empirical relationship between the solar flare X-ray energy and its associated CME mass \citep{2012ApJ...760....9A,2013ApJ...764..170D}, provides a second method to constrain CME mass.
\begin{equation}
M_{CME} = K_{M} E_{GOES}^{ \gamma} \mbox{  [g]}
\label{eq:5}
\end{equation}
Where $M_{CME}$ is the CME mass, $K_{M} = 2.7 \pm 1.2 \times 10^{-3}$ in cgs units, $\gamma = 0.63 \pm 0.04$ \citep{2012ApJ...760....9A}, and $E_{GOES}$ is the energy flare energy (erg) in the GOES (1-8 \AA) band.  
To apply this relation to the stellar case, a conversion from the GOES bandpass to the optical bandpass (u-band) is required,
\begin{equation}
\label{eq:empirical}
M_{CME} = K_{M} \left(\frac{f_{GOES}}{f_{u}}\right)^{\gamma} E_{u}^{\gamma} \mbox{  [g]}
\end{equation}
with $f_{u} = 0.11$ and $f_{GOES} = 0.06$ \citep{Osten2015} as the fractions of the bolometric flare energy designated to their respective frequency ranges of the optical bandpass and the GOES bandpass of 1-8 \AA .
As with the previous case, assuming that the solar relation extends to the stellar case, this provides a direct method for constraining the CME mass from observed optical flare energies.
There are large uncertainties in estimating mass from flare energies using equations (\ref{eq:5}) or (\ref{eq:empirical}). 
The uncertainties in $K_{M}$ and $\gamma$ together give about 1-2 orders of magnitudes spread.
Therefore, its success may be limited when examining events individually, but it should have more so when dealing with larger quantities of detections or while searching for average $M_{CME}$.  

In future experiments, a solar scaling test is to verify that equations \ref{eq:equipar} and \ref{eq:empirical} provide consistent results for $M_{CME}$.  
This test may bolster or refute the assumption that the solar scaling relations extend to the stellar case.
Assuming the test holds, equations \ref{eq:equipar} and \ref{eq:empirical} may be utilized to solve for flare energy as a function of CME velocity.
\begin{equation}
\label{eq:Combined}
E_{u} = \left[ \frac{\epsilon f_{u}}{2}K_{M} \left( \frac{f_{GOES}}{f_{u}}\right)^{\gamma} v^{2} \right]^{\frac{1}{1-\gamma}}
\end{equation}.  
This is now a single variable equation of $E_{u}(v)$.  
Simultaneous observations of radio and optical frequencies are required to evaluate this equation.  
The radio observations are able to provide drift rates of putative type II bursts produced by the CMEs, and thus velocity constraints.
Optical u-band observations are able to measure the associated flare energy.  

Equations  (\ref{eq:equipar}), (\ref{eq:empirical}), and (\ref{eq:Combined}) provide a theoretical answer to the question of 'What type of characteristics will this associated CME have for a flare of a certain energy?'
For this analysis, it is assumed that a detectable shock is caused by a CME traveling with a velocity slightly above the  Alfv\'{e}n speed generating an MHD shock in the traditional manner.
This method will provide a theoretical lower bound of flare energy required by a flare to produce a shock as it is just above the threshold of shock formation.    
Here the Alfv\'{e}n speed is represented with $v_{a}$ where B is the magnetic field strength and n is the local electron density. 
 
Assuming that the type II burst is slightly super-Alfv\'{e}nic ($1.2v_{a}$) and thus able to create a type II shock, the minimum energy required by the associated flare can be estimated.  
\citet{Zucca2014} suggests that solar type II shocks preferentially form near the solar surface ($\sim$ 1 - 2 R$_{\odot}$), and \citet{Susino2015} suggest that the entirety of the shock surface may be super-Alfv\'{e}nic at early phases (2 - 4 R$_{\odot}$).
 Therefore, it will be assumed initially that a shock will form at  $R = 2R_{\ast}$.   
Using typical numbers for M dwarfs at the base of the corona of $n_{0} = 3\times10^{10}$ cm$^{-3}$ \citep{2004A&A...427..667N} and B$_{0} =100$ G \citep{2007MNRAS.379.1075R}, the minimum flare energy becomes $E_{u} \approx 1.9\times10^{30}$ erg. 
This energy represents a moderately sized u-band flare.  
Table 3's bottom row lists the associated flare energies for type II bursts at various distances with the main trend of requiring less energy to generate shocks at further distances.
To compare this energy to that predicted by \citet{2006ApJ...650L.143Y} for where a one-to-one ratio occurs for flares and CMEs on the Sun, this value must be multiplied by $\left(\frac{f_{GOES}}{f_{u}}\right)$ or approximately 0.545.  
This brings the predicted flare energy to $E_{GOES} \approx 1.07\times10^{30}$ erg which is close to the predictions of \citet{2006ApJ...650L.143Y}. 
The minimum energy to shock is very sensitive to the initial conditions $n_{0}$ and $B_{0}$.
Lowering the magnetic field strength or increasing the number density all cause the minimum energy to shock to lower and vice versa for an increase. 

\begin{deluxetable}{c | c c c c c c c}	
\tabletypesize{\footnotesize}
\tablecaption{LOFAR Sensitivity}
\tablecolumns{8}
\tablewidth{0pt}
\tablehead{
\vspace{ -1.21cm}  
}
\startdata
	Distance in R$_{\ast}$ & 2 & 3 & 4 & 5 & 6 & 7 & 8 \\ \cline{2-8}
	Drift Rate [MHz/s] & 0.76 & 0.27 & 0.059 & 0.035 & 0.013 & 0.0046 & 0.0016\\
	Initial Frequency [MHz] & 347 & 164 & 53 & 37 & 17 & 8 & 4 \\	
	 E$_{u}$ [$10^{28}$ ergs] & 198 & 45.3 & 5 & 2.4 & 0.55 & 0.13 & 0.0029\\ 
\enddata
\tablecomments{This calculation assumes that the shocks initially form at the distance specified, therefore assuming the slowest possible shock (and lowest drift rate) which would generate a type II burst.  E$_{u}$ is the energy required of the associated flare to the CME in order to create a shock at that distance.  See text for more details.
}
\end{deluxetable}
\vspace{5mm}

\begin{figure}
\centering
\includegraphics[scale=1,clip=true, trim=2.5cm 8.75cm 0cm 7.5cm]{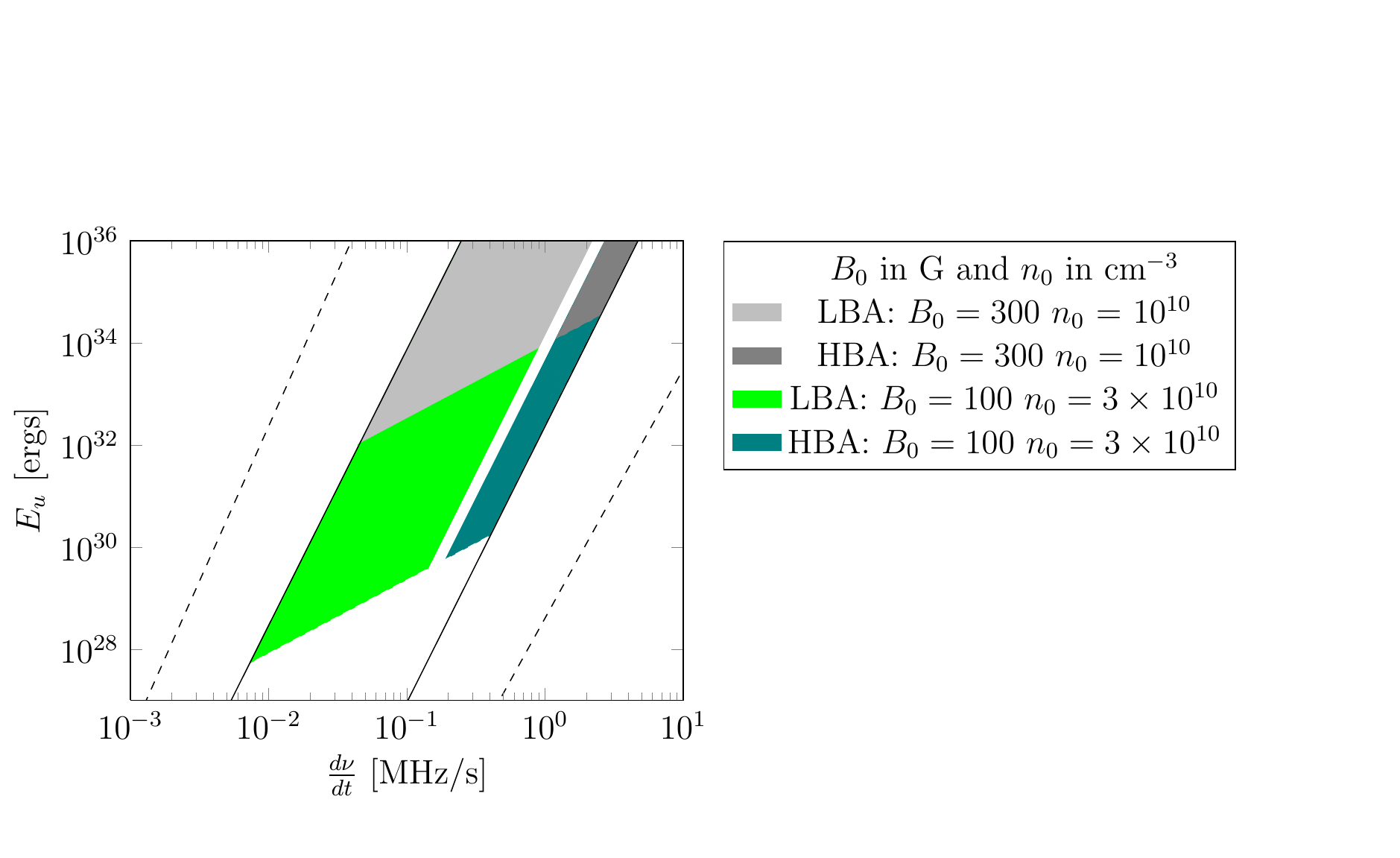}
\vspace{-1cm}
\caption{The associated flare energies as a function of drift rate for a potential type II shock.
The shaded region displays the parameter space that LOFAR is able to cover under various conditions.  
Light gray and green is used to represent LBA and dark gray and teal is used to represent HBA. 
Using a lower limit of 1.2$v_{a}$ for the minimum speed required to drive a shock ($cos\theta \approx $1) in LOFAR's frequency range defines the bottom edge.
The solid lines represent lines of constant frequency at 10 (left) and 190 (right) MHz.
The dashed lines correspond to the uncertainty of this region deriving from determining CME mass from flare energies.
The gray and dark gray shaded regions use values of B$_{0}=300$ G and $n_{0}=10^{10}$ cm$^{-3}$.  
The green and teal regions use the values B$_{0}=100$ G and $n_{0}=3\times10^{10}$ cm$^{-3}$.  
For clarity the green and teal regions continue into and cover the same areas as the gray and dark gray regions.
}
\label{fig:ShaddedPlot}
\end{figure}

By rewriting equation (\ref{eq:nudot}) to $v = \frac{2H_{0}}{cos\theta}\left(\frac{\dot{\nu}}{\nu}\right)$, equation (\ref{eq:Combined}) can be manipulated into the form:
\begin{equation}
\label{eq:combined2}
E_{u} = \left[ \frac{2H^{2}_{0}\epsilon f_{u}K_{M} }{cos^{2}\theta}\left( \frac{f_{GOES}}{f_{u}}\right)^{\gamma} \left(\frac{\dot{\nu}}{\nu}\right)^{2} \right]^{\frac{1}{1-\gamma}}
\end{equation}  
where $E_{u}(v)$ has become $E_{u}\left(\frac{\dot{\nu}}{\nu}\right)$. 
Figure \ref{fig:ShaddedPlot} displays equation (\ref{eq:combined2}) graphically for several specified frequency values assuming that $cos\theta \approx 1$.
Evaluating equation (\ref{eq:combined2}), with the parameters provided by \citet{2012ApJ...760....9A} for $K_{M}$ and $\gamma$, leads to the simplification $E_{u} \approx 4.75\times 10^{44} \left(\frac{\dot{\nu}}{\nu}\right)^{5.4}$.  

The two solid black lines represent lines of constant frequency 10 MHz (left) and 190 MHz (right).
The green/lightgray shaded region represents the parameter space where LOFAR's LBA system would be sensitive and the teal/dark gray region represents the parameter space where the HBA system would be sensitive under various stellar conditions.  
The figure uses a magnetic field strength of B$_{0}=100$ G and $n_{0}=3\times10^{10}$ cm$^{-3}$ at the base of the corona for the green and teal regions while it uses B$_{0}=300$ G and $n_{0}=10^{10}$ cm$^{-3}$ for the gray and dark gray regions.
For the LBA (green / light gray), the lowest green corners (along the 10 MHz line) correspond to $\dot{\nu} = 0.007$ MHz/s at energy $E_{u}\approx5\times10^{27}$erg, and the gray lowest corner $\dot{\nu} = 0.046$ MHz/s at energy $E_{u}\approx1.1\times10^{32}$erg.  
For the HBA (teal / dark gray), the lowest teal corners (which would follow a 110 MHz line) correspond to $\dot{\nu} = 0.19 MHz/s$ at energy $E_{u}\approx5.5\times10^{29}$erg, and the dark gray lowest corner $\dot{\nu} = 1.2$ MH/s at energy $E_{u}\approx1.2\times10^{34}$erg.  
As an example to read this plot, if a 10 MHz burst was observed to have a frequency drift rate of $3.5 \times 10^{-2}$ MHz/s, then the energy of a flare that would have been associated to that drift rate would be about $10^{32}$ ergs.  

The bottom edge of each shaded region in Figure \ref{fig:ShaddedPlot} is defined as the energy of a flare that would be associated with a shock that is moving at 1.2$v_{a} \left( \propto \frac{B}{\sqrt{n}} \right)$.
This is just above the minimum speed required by a CME to generate an MHD shock.
This is a physically based constraint and not a technical limitation for observation.
Both increasing the number density or lowering the magnetic field strength will push the left most edge lower down the lines of constant frequency.  
There is no upper boundary as it is possible to have arbitrarily energetic shocks that would be observable.  
For clarity this means that the green and teal regions will extend into the gray/dark gray regions and beyond.  

Figure \ref{fig:ShaddedPlot} is intended to emphasize that the observable flare energies will be constrained to a range of drift rates and can change dramatically based on the conditions of the star. 
The green and teal regions represent the theoretical sensitivity for a star similar to YZ CMi, while the gray and dark gray regions represent conditions that differ by 1/3 the density and 3 times the magnetic field strength. 
Using the flare energy shown by \citet{2006ApJ...650L.143Y} to have a one-to-one correspondence between flares and CMEs (10$^{31}$ ergs), the sensitivity to stellar parameters can be shown.  
A YZ CMi type star with a type II burst emitting in LOFAR's observing frequency range would be fully covering the parameter space (green and teal regions been the lines of constant frequency of 10 and190 MHz for a constant energy of $10^{31}$ ergs).

Alternatively, a star similar to YZ CMi but with 1/3 the base coronal density and 3 times the magnetic field strength with a type II burst emitting in LOFAR's observing frequencies would have a signal that would be completely outside of the parameter space (gray and dark gray regions) that LOFAR would be able to cover.  
Here the gray and dark gray regions do not extend down to the $10^{31}$ ergs energy region between the lines of constant frequency 10 and 190 MHz and thus would be unable to detect CMEs corresponding to a flare with this energy.   

A star with low densities and strong magnetic fields may not produce an observable shock with a flare of certain energy while a star with high density and weak magnetic fields would produce a detectable shock off an identical flare.
Preferential stars to observe for shocks are those stars with lower magnetic field strengths and higher coronal electron densities while keeping high flare rates.    
This predicts detectable drift rates of $\sim10^{-1}$ MHz/s.  

The two dashed lines in Figure (\ref{fig:ShaddedPlot}) are representative of the uncertainty in the parameter space induced by the uncertainties of $K_{M}$ and $\gamma$ from \citet{2012ApJ...760....9A}.
The large spread in \citet {2012ApJ...760....9A} leading to equation (\ref{eq:5}) has the majority of its impact on equation (\ref{eq:combined2}) via the uncertainties in the power $\gamma$ despite the fact that it has a lower fractional-uncertainty $\left(\frac{\delta\gamma}{\gamma}\right)$ than $K_{M}$.  
This creates the potential for several orders of magnitude uncertainty in either directions about any specific measurement.  
Since the empirical relations in \citet{2012ApJ...760....9A} and \citet{2013ApJ...764..170D} provide consistent values for CME mass, even given wide spread, a large number of dual measurements are needed to test accurately the validity of solar-stellar scaling.

\section{Conclusion}

The observations presented in Table 1 have yielded no plausible detections of drifting low-frequency (type II) radio bursts on YZ CMi and thus no CME detections either. 
Assuming the rate of shocks is a lower limit to the rate at which CMEs occur, no detections in a total of 15 hours of observation places a limit of $\nu_{type II} <  0.0667$ shocks/hr $ \leq \nu_{CME}$ for YZ CMi due to the stochastic nature of the events and limits of observational sensitivity.  
This rate is for the frequency range between 30-160 MHz, at the sensitivities stated above, assuming the rate of shocks is less than or equal to the rate of CMEs.
Due to the lack of detections, and the exploratory nature of this paper, no rigorous constraints can yet be placed on whether flares on M dwarfs have CMEs.

A likely contributor to the lack of detections is that CMEs producing a type II shock in the correct parameter space also need to have a sufficiently high brightness temperature and/or large enough angular size to produce a flux density detectable by LOFAR.  
More sensitive observations (with more stations) and longer observation periods would increase the likelihood of detections. 

Based on Section \ref{MethodologyForConnection}, simultaneous observations between radio and u-band would allow for a deeper investigation of M dwarf CMEs.  
Stars with weaker magnetic field strengths and higher electron densities of the corona while maintaining high flare rates are preferred for type II burst detections.  
Weaker flares are related to slower CMEs via equation (\ref{eq:Combined}).  
This means that the observable drift rates are slower due to the shock forming further away from the star.  
Generating high drift rate shocks require shock formation closer to the star in its higher density regions and require more energy reducing the frequency of occurrence.  
Simultaneous observation provides direct flare measurements which would allow for CME property predictions that could be tested against CME detections if found.  

Given the assumptions and limitations of this current dataset, rigorous limits can not yet be placed on the occurrence and properties of stellar CME's.
The lack of any energy, mass, or velocity constraints also prevents any further constraints to modeling CMEs impact on nearby exoplanets. 
New low frequency instruments like LOFAR, Murchison Widefield Array (80 - 300 MHz), Karl G. Janskey Very Large Array's p-band (230-470 MHz), and the future Square Kilometer Array are providing new opportunities to advance this field and further develop this line of discovery. 

\acknowledgements
MKC and RAO acknowledges funding support from NSF AST-1412525 for the project on which this paper is based.  
LOFAR, the Low Frequency Array designed and constructed by ASTRON, has facilities in several countries, that are owned by various parties (each with their own funding sources), and that are collectively operated by the International LOFAR Telescope (ILT) foundation under a joint scientific policy.
SC acknowledges financial support from the UnivEarthS Labex program of Sorbonne Paris Cit\'e (ANR-10-LABX-0023 and ANR-11-IDEX-0005-02).  
JvL acknowledges funding from the European Research Council under the European Union's Seventh Framework Programme (FP/2007-2013) / ERC Grant Agreement n.617199.
MKC and RAO thank Tim Bastian for his scientific consultation.

\end{document}